\documentstyle[aps,epsf,rotate,preprint]{revtex}

\begin{document}

\draft

\title{Behavioral-Independent Features of Complex Heartbeat Dynamics}

\author{ Lu\'{\i}s~A.~Nunes
	 Amaral$^{1,2,*}$, Plamen
	 Ch. Ivanov$^{1,2}$, Naoko Aoyagi$^3$, Ichiro Hidaka$^3$, \\ Shinji
	 Tomono$^3$, Ary L. Goldberger$^2$, H. Eugene Stanley$^1$, and
	 Yoshiharu Yamamoto$^3$ }

\address{ $^1$Center for Polymer Studies and Department of Physics,
	     Boston University, Boston, MA 02215 \\
	  $^2$Cardiovascular Division,  Beth Israel Deaconess Medical 
	     Center, 
	     Harvard Medical School, Boston, MA 02215 \\
	  $^3$Educational Physiology Laboratory, Graduate School of
	     Education, 
	     University of Tokyo, Bunkyo-ku, Tokyo 113-0033, Japan
	}

\maketitle

\begin{abstract}

We test whether the complexity of cardiac interbeat interval time
series is simply a consequence of the wide range of scales
characterizing human behavior, especially physical activity, by
analyzing data taken from healthy adult subjects under three
conditions with controls: (i) a ``constant routine'' protocol where
physical activity and postural changes are kept to a minimum, (ii)
sympathetic blockade, and (iii) parasympathetic blockade.  We find
that when fluctuations in physical activity and other behavioral
modifiers are minimized, a remarkable level of complexity of heartbeat
dynamics remains, while for neuroautonomic blockade the
multifractal complexity decreases.

\end{abstract}

\noindent



Healthy free-running physiologic systems have complex self-regulating
mechanisms which process inputs with a broad range of characteristics
\cite{Shlesinger}, and may generate signals that have scale-invariant
dynamics \cite{Musha}.  Many physiologic time series are extremely
``patchy'' and nonstationary, fluctuating in an irregular and complex
manner.  This observation suggests that some physiologic signals are
sufficiently inhomogeneous that a single fractal exponent may not be
sufficient to characterize them.

Time series of healthy human interbeat intervals belong to a special
class of complex signals that display multifractal properties
\cite{Ivanov99}.  Multifractal signals---such as those generated by
binomial multiplicative processes or turbulent fluctuations--- can be
decomposed into many subsets characterized by different {\it local\/}
Hurst \cite{Vicsek} exponents $h$, which quantify the local singular
behavior and thus relate to the local fractal properties of the time
series \cite{Vicsek,Barabasi}.  The statistical
properties of the different subsets characterized by the different
exponent values of $h$ are quantified by the function $D(h)$.  Here
$D(h_o)$ is the fractal dimension of the subset of the original time
series characterized by the local Hurst exponent $h_o$
\cite{Vicsek,Barabasi}.  For heart rate time series
from healthy individuals, the function $D(h)$ is ``broad'' (implying
multifractality), but ``narrow'' (implying monofractality) for
subjects with heart failure \cite{Ivanov99}, a life-threatening
condition.

An intriguing question, with implications for basic signalling and
feedback mechanisms, is what gives rise to multifractality in healthy
human heartbeat dynamics?  Two distinct possibilities can be
considered.  The first is that the observed multifractality is
primarily a consequence of the response of neuroautonomic control
mechanisms to activity-related fractal stimuli \cite{Musha}.  If this
were the case, then in the absence of such correlated inputs the
heartbeat dynamics would not generate such a heterogeneous
multifractal output.  The second is that the neuroautonomic control
mechanisms ---in the presence of even weak external noise--- {\it
endogenously\/} generate multifractal dynamics.  Here, we present
evidence from three new experiments which supports the latter
possibility.


The procedure to calculate the values of $h$ and their corresponding
fractal dimensions has been described elsewhere
\cite{Ivanov99,Daubechies88,Muzy,Details_1}.  We calculate the
experimental $\tau(q)$, which is related to $D(h)$ through a Legendre
transform \cite{Vicsek,Details_2,Physionet},
%
\begin{equation}
D(h) = q~\frac{d \tau(q)}{d q} - \tau(q)\,.
\label{e-fa}
\end{equation}

\begin{figure}
\narrowtext
\centerline{
\epsfysize=0.5\columnwidth{{\epsfbox{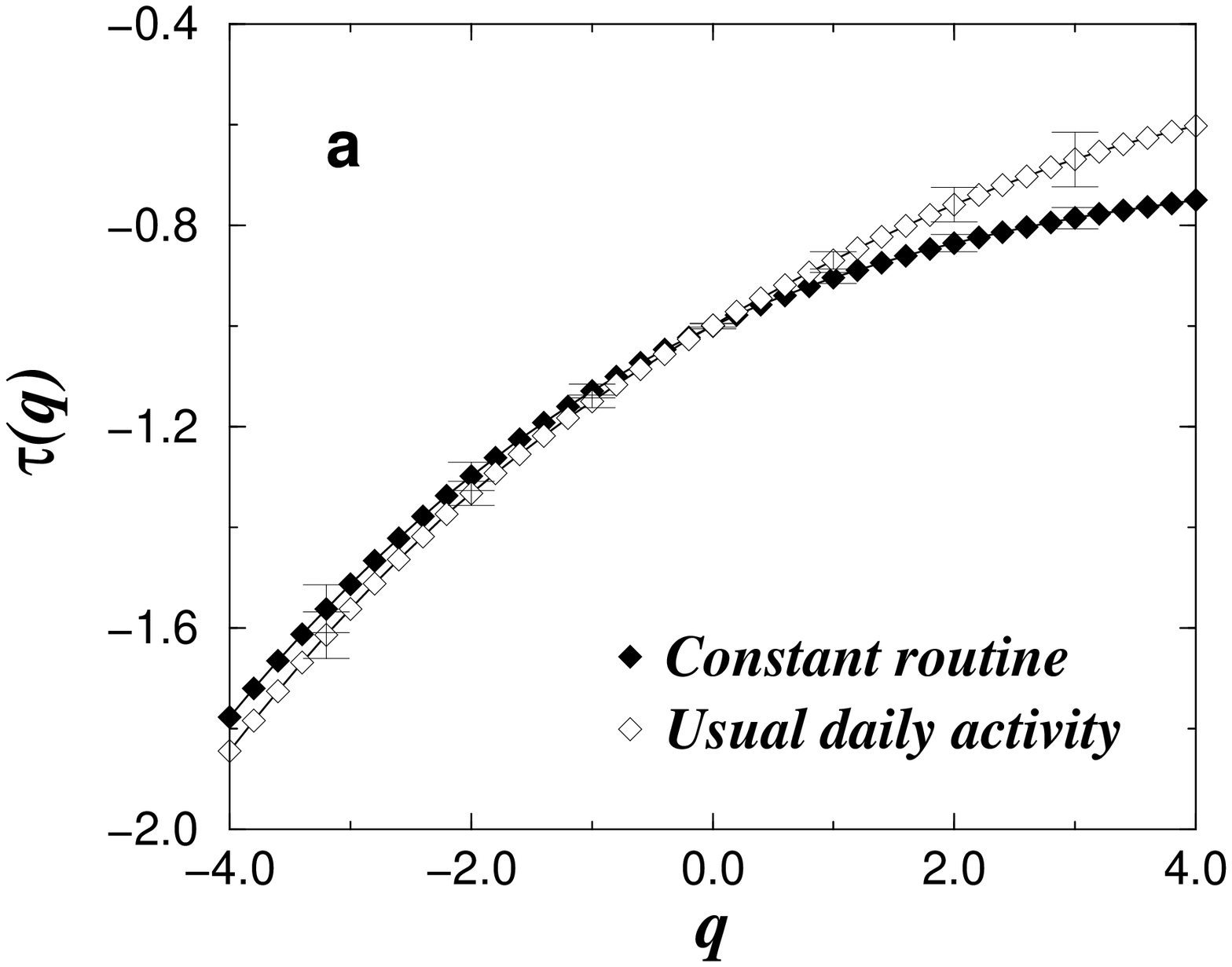}}}
}
\centerline{
\epsfysize=0.5\columnwidth{{\epsfbox{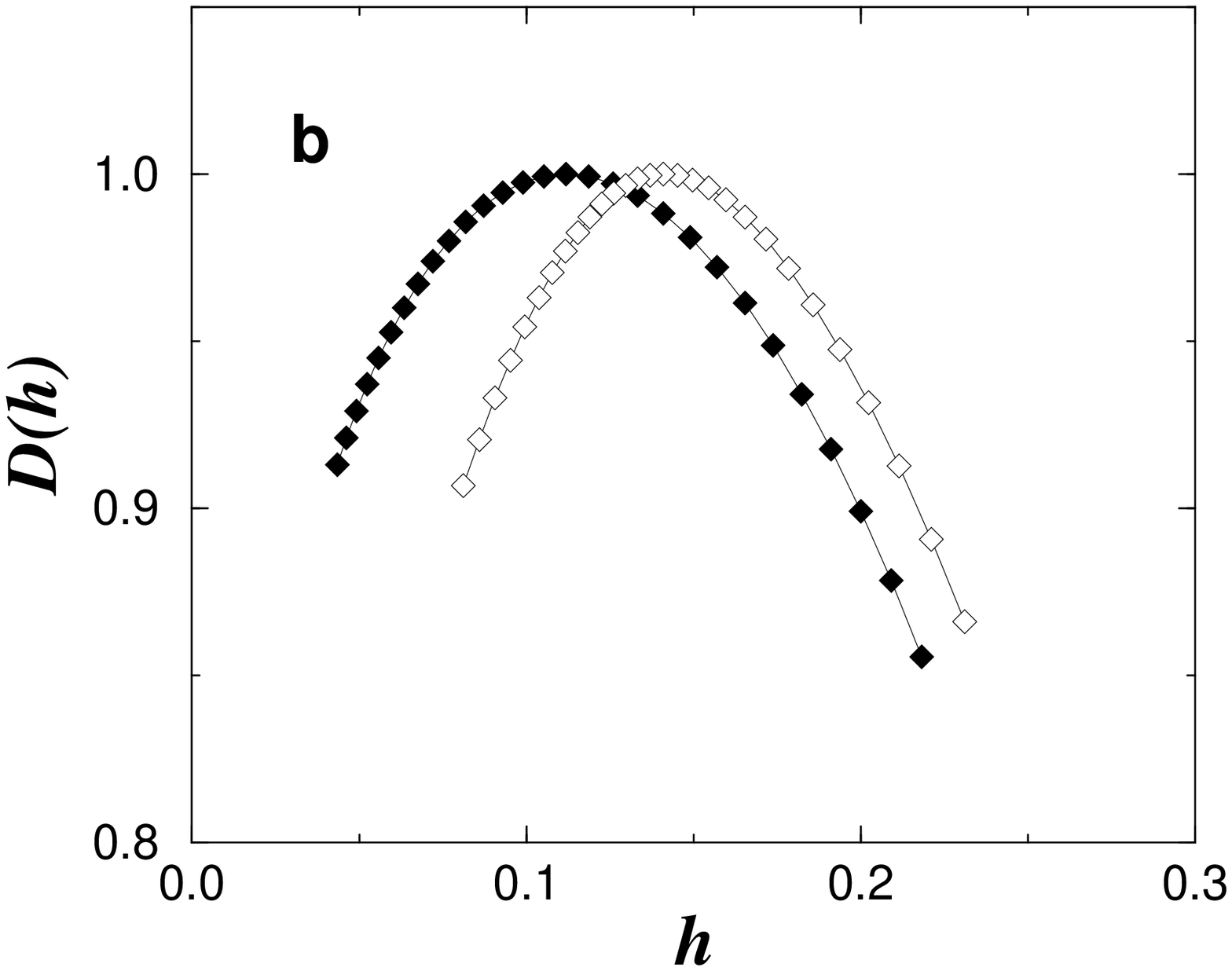}}}
}
\vspace*{0.cm}
\caption{ Constant routine study. The average heart beat interval for
constant routine is 0.951 s (0.820 s for controls) and the average
standard deviation is 0.117 s (0.129 s for controls).
(a) Multifractal spectra $\tau(q)$ for constant routine and control
(usual daily activity) protocols ($n=6$).  In this and following
figures, the error bars indicate the standard error of the group
average $\tau(q)$.  The two curves have nearly identical curvature but
appear to be slightly rotated around a vertical axis going through
$q=0$, which suggests that there are no major differences in the
multifractal properties of control and constant routine groups.
(b) Singularity spectra $D(h)$ for the two groups.  $D(h)$, which is
obtained as the Legendre transform of $\tau(q)$, measures the fractal
dimension of the subsets of the signal characterized by local Hurst
exponents $h$.  Note that the two curves have nearly identical widths
indicating a similar degree of multifractality.  This result is
consistent with the possibility that the activities of daily living do
not account for the multifractal complexity of heart rate dynamics. }
\label{f-cr}
\end{figure}

We first analyze datasets from six healthy, non-smoking male subjects
\cite{Tokyo} (ages: 21--30~yr).  We obtained two datasets per subject,
the first under constant routine conditions, and the second under
usual daily activity conditions \cite{Constant_rout,na00}.
Figure~\ref{f-cr}a displays the average multifractal spectra $\tau(q)$
for the six subjects under both regimens.  The nonlinearity of
$\tau(q)$ does not appear altered by constant routine conditions.
Indeed, our analysis indicates that major reductions in external
stimulation and physical activity do not reduce the multifractal
properties of healthy cardiac dynamics, supporting the hypothesis that
the multifractality in healthy heartbeat dynamics is endogenous to the
neuroautonomic regulation of the heart rate \cite{Meesmann,Fortrat}.

\begin{figure}
\narrowtext
\centerline{
\epsfysize=0.5\columnwidth{{\epsfbox{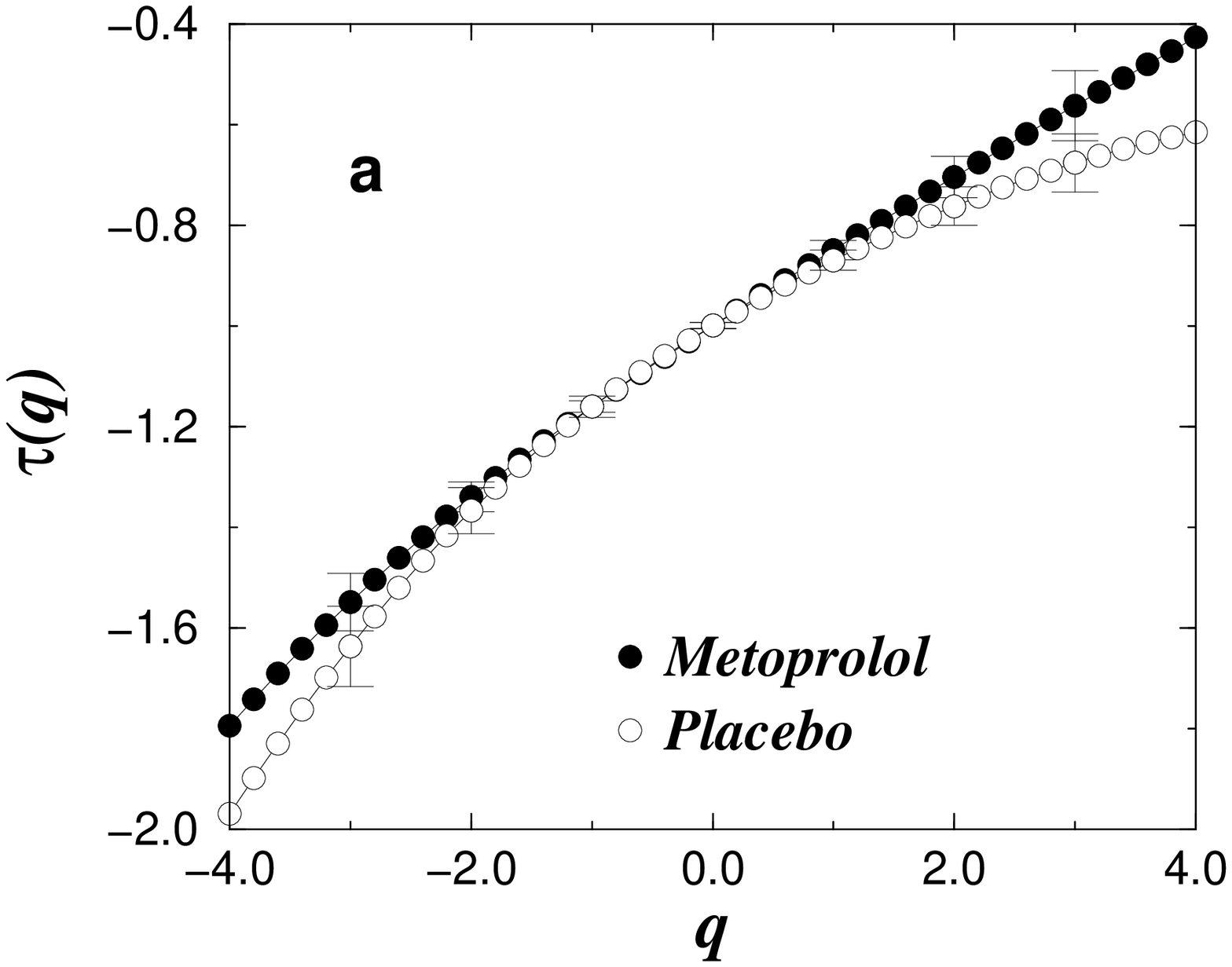}}}
}
\centerline{
\epsfysize=0.5\columnwidth{{\epsfbox{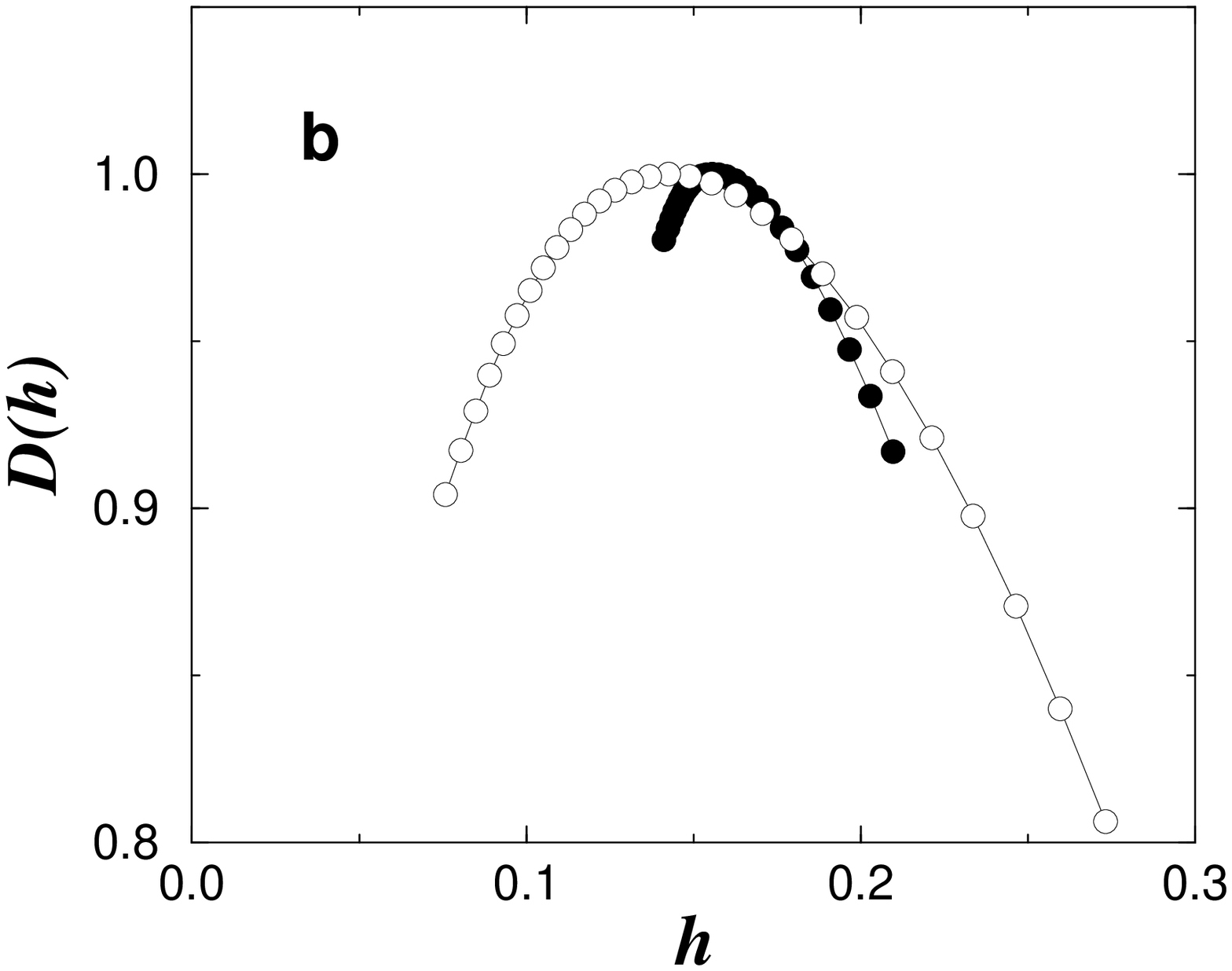}}}
}
\vspace*{0.cm}
\caption{ Sympathetic blockade study.  The average heart beat interval
for beta-blocker intake is 0.885 s (0.750 s for placebo intake) and
the average standard deviation is 0.115 s (0.092 s for placebo
intake).
(a) Group average $\tau(q)$ for datasets ($n=8$) during sympathetic
(beta) blockade with metoprolol \protect\cite{McMurray}.  Note that
the curve for the beta blocker administration remains above the
placebo curves for all $q$, indicating a more marked curvature
---i.e., {\it nonlinearity}--- of $\tau(q)$ for the control group.
(b) Singularity spectra $D(h)$ for the two groups.  The singularity
spectrum is obtained by a Legendre transform of the multifractal
spectrum.  Note that the peaks for the two curves appear to be at
somewhat different positions.  However, our analysis indicates that
the variation from group to group and subject to subject is wider for
the position of the peak than for its width, so that no change in peak
position can be inferred.  Heartbeat dynamics during sympathetic
blockade display a change in the singularity spectrum, namely
decreased multifractality as evidenced by the narrower distribution
$D(h)$. The ``weaker'' nonlinearity of $\tau(q)$ during sympathetic
blockade indicates less pronounced multifractality, suggesting that
suppression of sympathetic control decreases the multifractal
complexity of cardiac dynamics. }
\label{f-bl}
\end{figure}

To test the possibility that the multifractality in healthy heartbeat
dynamics is related to neuroautonomic control, we analyze data from
six subjects \cite{Tokyo} (4 male, 2 female, ages: 21--34~yr) who were
administered a beta-blocking drug \cite{McMurray} which reduces
sympathetic control.  We analyzed eight datasets from the six subjects
from the the second and/or third day of beta-blocker administration
\cite{na00,Beta-blocker}.  As a control, we also analyzed eight
datasets for the same subjects but for the second and/or third day of
placebo administration \cite{Beta-blocker}.  Figure~\ref{f-bl}a shows
the multifractal spectra for the two groups. The curve for the group
corresponding to the administration of the the beta-blocker drug is
more linear than that for the control group.  This result is
consistent with decreased multifractality due to the suppression of
sympathetic control (Fig.~\ref{f-bl}b).

\begin{figure}
\narrowtext
\centerline{
\epsfysize=0.5\columnwidth{{\epsfbox{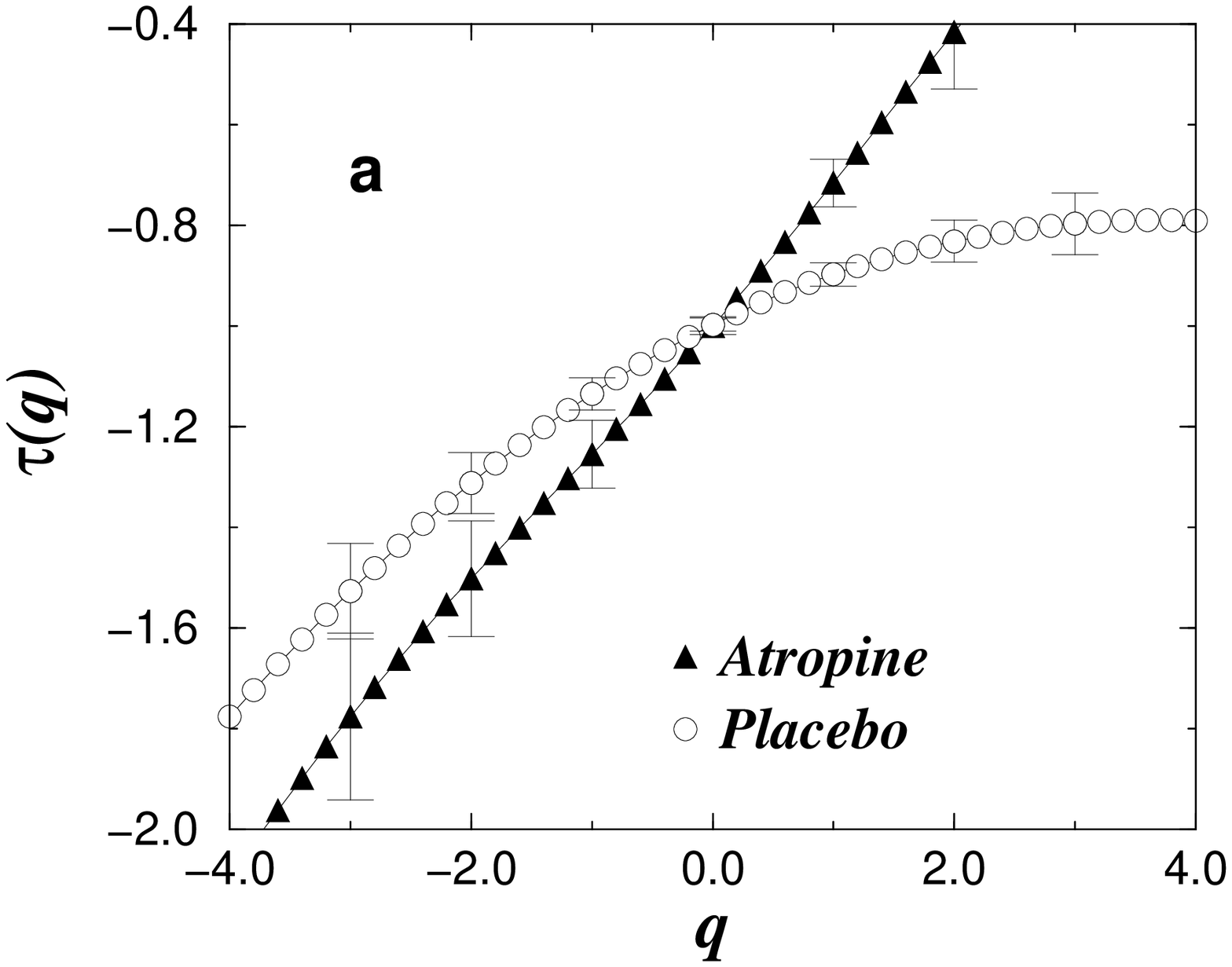}}}
}
\centerline{
\epsfysize=0.5\columnwidth{{\epsfbox{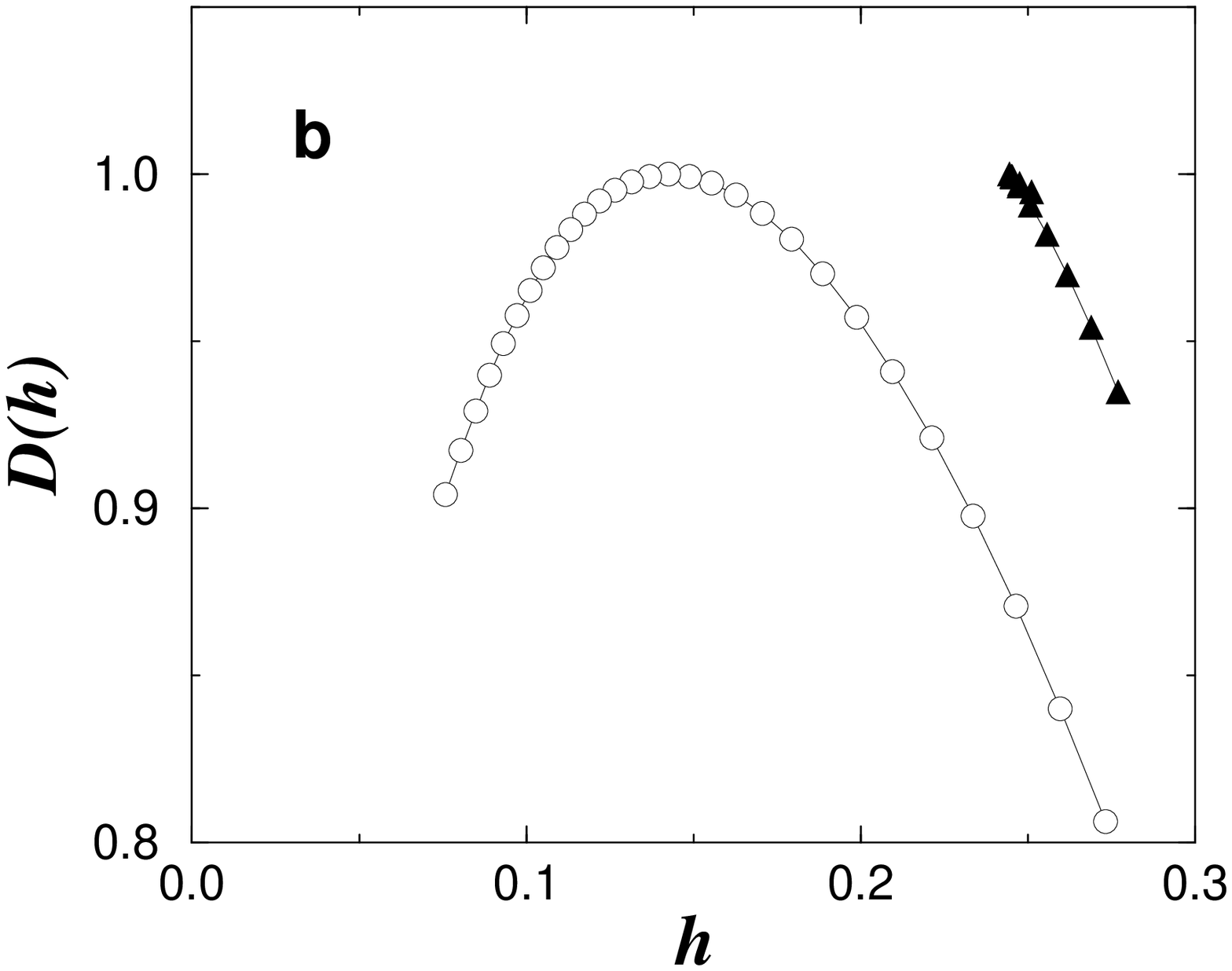}}}
}
\vspace*{0.cm}
\caption{ Parasympathetic blockade study. The average heart beat
interval for atropine intake is 0.703 s and the average standard
deviation is 0.078 s.
(a) Group average $\tau(q)$ for datasets obtained during
parasympathetic (vagal) blockade with atropine ($n=6$).  Because of
the potential adverse effects associated with very prolonged
parasympathetic blockade, these datasets are shorter than the others,
consisting of only about 6,000 interbeat intervals. As a control, we
analyze the first 6,000 data points from the subjects being
administered the placebo in the sympathetic blockade experiments. Our
analysis suggests (i) that the dynamics become monofractal under
parasympathetic blockade ---note that $\tau(q)$ becomes nearly
linear--- and (ii) that the typical Hurst exponent increases towards
less anti-correlated values as previously observed for severe heart
failure ($h_{HF} \approx 0.25$) \protect\cite{Musha} ---note the
increase in the slope for $q$ close to zero which is closely related
to the single exponent obtained by a standard (mono)fractal analysis
\protect\cite{Musha}.
(b) Singularity spectra $D(h)$ for the two groups. The singularity
spectrum is obtained by a Legendre transform of the multifractal
spectrum. The figure shows that the heart rate dynamics after
parasympathetic blockade becomes nearly monofractal.
}
\label{f-at}
\end{figure}

As a further test, we also analyze the multifractal properties of the
heartbeat dynamics of healthy individuals who were administered
atropine \cite{atropine} which suppresses parasympathetic control of the
heartbeat.  We analyze six datasets from six different healthy males
\cite{Tokyo} (ages: 21--26 yr).  As a control, we utilize datasets
from subjects in the beta-blockade experiment after administration of
the placebo.  Figure~\ref{f-at}a shows the multifractal spectra for
the two groups. The curve for the group under parasympathetic blockade
is nearly linear ---indicating a marked loss of multifractality---
even more apparent than with sympathetic blockade (Fig.~\ref{f-at}b).
These results are consistent with the possibility that multifractality
in healthy heartbeat dynamics may arise, at least in part, from the
interplay between the two branches of the neuroautonomic system.


The major finding of this study is the strong evidence supporting the idea
that (multi)fractality in heartbeat dynamics is related to intrinsic
properties of the control mechanisms, and is not simply due to changes in
external stimulation, degree of physical activity or other apparent
behavioral modifiers---such as postural changes, food intake and sleep-phase
transitions (See Table I).  Understanding how the interaction of
neuroautonomic, and possibly other, control mechanisms generates the complex
multiscale dynamics of the heartbeat will be a major challenge to future
efforts to model ``real-world'' signalling mechanisms \cite{Marshall00}.

Our results are also of note for a number of other reasons.  First, as shown
in Figs.~\ref{f-cr}-\ref{f-at}, the singularity spectrum $D(h)$ during
sympathetic blockade has a narrower range of allowed values of $h$ than the
singularity spectra for the control groups.  However, the position of the
peak in $D(h)$ during sympathetic blockade is not substantially modified from
its position for the same subjects when given a placebo.  This suggests that
sympathetic blockade may not have a major effect on the linear correlations
in the dynamics, that is, it does not change the average Hurst exponent
substantially \cite{Yamamoto94}.

Second, we find that during parasympathetic blockade there is a marked loss
of multifractality (see Fig.~\ref{f-at}b), much as occurs for patients with
severe heart failure \cite{Ivanov99}.  Indeed, as with heart failure, the
peak of the singularity spectra is located to the right of the healthy
control group, indicating weaker anticorrelations \cite{Musha}.  This finding
is consistent with the hypothesis that both the monofractality and ``weaker''
anti-correlations for heart failure dynamics may be related, at least in
part, to impaired parasympathetic control in congestive heart failure
patients, in agreement with recent studies \cite{Yamamoto95}.

Third, our finding of the impact of neuroautonomic control on the
multifractal properties of heart rate variability during waking hours raises
the question of how transitions during sleep might affect these properties.
The present study using a constant routine protocol was not designed to
elucidate this intriguing possibility.  Recent reports of differences in
heartbeat scaling exponents---related to two-point correlations---between
daytime and nighttime hours \cite{Ivanov} and also during different sleep
stages \cite{Bunde-sleep} support the need for future investigation of
multifractal properties in different physiologic states.

Finally, we note that for many of the systems which generate
multifractal signals there are no mathematical equations describing
the dynamics, and even for those for which such equations exist, their
analytical solution is not feasible.  Thus, the understanding and
modeling of the intrinsic control mechanisms for heart rate may offer
new opportunities to explore multifractal dynamics in the natural
sciences \cite{model}.

We thank Y. Ashkenazy, L. Glass, J. M. Hausdorff, S. Havlin, S. Mossa,
C.-K. Peng, V. Schulte-Frohlinde, and Z. Struzik for helpful
discussions and suggestions.  This work was supported by grants from
NIH/NCRR (P41 RR13622) \cite{Physionet}, the Fetzer Institute, the
Mathers Charitable Foundation, the Centers for Disease Control
(H75/CCU119124), the Japan Space Foundation, and a Monbusho
Grant-in-Aid for Scientific Research.


\vspace*{1cm}
\noindent
$^{*}$~Email:~amaral@buphy.bu.edu; \\
WWW: http://polymer.bu.edu/$\sim$amaral/Heart.html.

%
%
%
%

%
%
\begin{table}
\narrowtext
\caption{ Width and peak position for $D(h)$ spectra for the
different protocols studied.
}
\vspace*{0.1cm}
\begin{tabular}[t]{l c c}
{\bf Protocol} & {\bf Width} & {\bf Peak} \\
\hline
Usual daily activity 	 & $0.16 \pm 0.04$ & $0.14 \pm 0.04$ \\
Constant routine     	 & $0.18 \pm 0.04$ & $0.11 \pm 0.04$ \\
\hline
Placebo 		 & $0.20 \pm 0.04$ & $0.14 \pm 0.04$ \\
Sympathetic blockade 	 & $0.08 \pm 0.04$ & $0.16 \pm 0.04$ \\
Parasympathetic blockade & $0.03 \pm 0.03$ & $0.24 \pm 0.02$ \\
\end{tabular}
\end{table}
%
%



\begin{thebibliography}{99}
\bibitem{Tokyo} Each subject gave informed consent to participate in
this study after the test protocol was fully described.  This study
was reviewed and approved by the ethics committee of the Graduate
School of Education of The University of Tokyo.
\bibitem{Fortrat} Our results do not preclude the possibility that
even very slight movements may affect heart rate dynamics and
contribute to the observed multifractal properties.  However, it is
not feasible to prevent all motion in a controlled way.  Further,
attempts to impose such restraints would be stressful and alter
neuroautonomic tone.  Moreover, many small movements, such as those
required for balance, are of an involuntary nature, which would
support our conclusion that voluntary physical activity is not
responsible for the observed complexity of heart rate variability.
\bibitem{Beta-blocker} In the sympathetic blockade study, 6 subjects
were randomly assigned into two groups and continuously monitored for
seven days. The first group of 3 subjects took a long-acting
$\beta_1$-selective adrenergic antagonist, metoprolol tartrate, in
single-blind fashion, 120 mg once daily in the morning for the first
three days.  After a wash-out day without any medication, the subjects
were given a placebo drug for the last three days. The second group of
3 subjects took the placebo first, followed by the metoprolol.
\bibitem{atropine} In the parasympathetic blockade study, data were
collected for 60~min under sustained suppression of the vagal
influence on heart rate by a primed (0.75~mg) constant intravenous
infusion (0.009~mg/min) of atropine sulfate. The infusion rate was
determined so that steady-state increases in heart rate of about
20~beats/min were achieved. The subjects were examined on a tilt table
while the tilt angle was changed every 10--15~min to investigate
cardiovascular responses to the orthostatic challenge, the original
purpose of the experiment. These transient changes in posture
introduce an additional nonstationarity in the data which, however,
can be eliminated by the application of wavelet analysis. The smaller
number of data points ($\approx$ 6,000 beats) obtained in this
experiment is nonetheless sufficient, as the largest wavelet scale we
use is only 400 beats. Heart rate was measured continuously from a
surface ECG by use of standard bipolar leads with an ECG device
(AT-601G, Nihon-Koden). The analog output of the ECG device was
differentiated to yield a train of rectangular impulses corresponding
to the QRS spikes, and the impulse train was processed at a sampling
frequency of 1,000~Hz to yield beat-to-beat RR intervals.
\end{thebibliography}
\end{document}